Article

# Analysing the Influence of Macroeconomic Factors on Credit Risk in the UK Banking Sector

Hemlata Sharma [1], Aparna Andhalkar [1], Oluwaseun Ajao [2] and Bayode Ogunleye [3,*]

1 Department of Computing, Sheffield Hallam University, Sheffield S1 2NU, UK
2 Department of Computing & Mathematics, Manchester Metropolitan University, Manchester M1 5GD, UK
3 Department of Computing & Mathematics, University of Brighton, Brighton BN2 4GJ, UK
* Correspondence: b.ogunleye@brighton.ac.uk

**Abstract:** Macroeconomic factors have a critical impact on banking credit risk, which cannot be directly controlled by banks, and therefore, there is a need for an early credit risk warning system based on the macroeconomy. By comparing different predictive models (traditional statistical and machine learning algorithms), this study aims to examine the macroeconomic determinants' impact on the UK banking credit risk and assess the most accurate credit risk estimate using predictive analytics. This study found that the variance-based multi-split decision tree algorithm is the most precise predictive model with interpretable, reliable, and robust results. Our model performance achieved 95% accuracy and evidenced that *unemployment* and *inflation rate* are significant credit risk predictors in the UK banking context. Our findings provided valuable insights such as a positive association between credit risk and inflation, the unemployment rate, and national savings, as well as a negative relationship between credit risk and national debt, total trade deficit, and national income. In addition, we empirically showed the relationship between national savings and non-performing loans, thus proving the "paradox of thrift". These findings benefit the credit risk management team in monitoring the macroeconomic factors' thresholds and implementing critical reforms to mitigate credit risk.

**Keywords:** banking; credit risk; decision tree; macroeconomic determinants; machine learning; predictive analysis

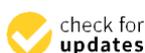

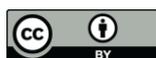

## 1. Introduction

Banks serve as the bedrock of the global financial ecosystem as they facilitate actual financial transactions with the movement of money amongst individuals, businesses, and governments, both domestically and internationally. Non-payment of debts causes significant losses to banks and is referred to as credit risk or a non-performing loan (NPL) [1]. According to an empirical study, NPLs are a significant and key indicator of credit risk; they are used as a precursor to the beginning of a financial crisis [2]. Credit risk has been considered as a critical risk by the International Monetary Fund (IMF) to the UK banking sector; therefore, consistent increases in NPLs are dangerous for banks [3]. Over the years, financial crises have had a substantial impact on banking stability. The 2008 global crisis revealed the interwoven nature of banking and macroeconomic indicators such as unemployment, inflation, etc. Also, it showed that a negative shift in macroeconomic indicators such as the unemployment rate, inflation, GDP, etc., initiates a vicious cycle, causing financial stress in the ecosystem [4]. Since COVID-19, global financial circumstances have deteriorated and are becoming worse because of the Russia–Ukraine war. This geopolitical uncertainty is causing inflation with energy bill shocks and an alarmingly uncontrollable global (debt) credit risk [5].

This study is focused on the UK, a country experiencing stagflation (increasing inflation and slowed down economic growth) with a forecast of a potential recession in 2023.

The Bank of England (BOE) warned UK banks to closely monitor credit risk and implement an early warning system which can emphasize the trajectory of macroeconomic indicators and find out the possibility of recession [5]. The above-mentioned discussions require the implementation of a decisive preventive action plan by UK banks to reduce the macro-economically driven credit risk, which can be achieved with advanced analytical insights. Credit risk multivariate and predictive models have been researched theoretically and empirically [6]; however, studies that consider and emphasise macroeconomic variables are limited. Thus, this paper aims to investigate the UK's macroeconomic determinants of its banking credit risk from 2005 to 2021. To this end, four research questions (RQs) were defined, targeting the beneficiary of the credit risk management team.

1. How did the UK's macroeconomic factors and credit risk change over the time from 2005 to 2021?
2. What was the effect of macroeconomic factors on credit risk from 2005 to 2021?
3. How are macroeconomic factors and banking credit risk related?
4. Which machine learning (ML) model can outperform conventional regression models for credit risk prediction?

The research scope covers different aspects of advanced analytics as a practical solution that facilitates decision intelligence, using UK banking credit risk data and macroeconomic variables. This study benefits stakeholders (credit risk managers and teams, risk analysts, auditors, senior management) in the banking industry to inform their decision-making processes.

To summarize this paper, our research answers the RQs in five sections. Section 2 discusses research gaps with proposed solutions in a literature review. Section 3 presents the methodology, and the findings are depicted in Section 4. Lastly, Section 5 concludes the research findings with a few recommendations, constraints, and the future scope.

**2. Related Work**

This section is a review of the existing literature that covers the business domain, the technological aspect of analytical solution in five themes, and recognizes research gaps.

*2.1. Theme 1: Credit Risk Definition, Indicators, and Implications*

Risks in banking can have multiple interpretations, like "prospective loss" triggered by adverse circumstances [7] or an "uncertainty about future outcome" [8]. As both interpretations are significant, presenting either solution will not suffice the purpose [9]. Therefore, this research focuses on both aspects. Bank credit risk is represented by different indicators such as NPLs [10,11] non-performing assets (NPAs), and the ratio of capital adequacy (CAR) [12]. The majority of researchers have empirically established NPLs as the primary indicator of banking credit risk [2], as NPLs can be used to calculate NPAs and CAR. It has been observed that there is a scarcity of UK-focused NPL research that covers the trend for a longer period of time to showcase a clear picture of trend.

*2.2. Theme 2: Selection of Credit Risk (NPL) Determinant Types*

Existing research divides NPL determinants into three groups—bank-specific, industry-related, and macroeconomic factors. As part of credit risk management, banks and the country's central bank actively govern bank-specific factors which affect NPLs [13]. On the other hand, industry-related factors like regulatory institutions also influence NPLs [14]. Macroeconomic factors define economic conditions at the global and/or national levels that are not directly within the control of banks. The core thesis is that the ability of loan repayment is influenced by the economic cycle of a nation or the globe, and this thesis is supported by several research studies. As a result, banks emphasize them for efficient credit risk management and economists include them when formulating policies [15–17]. Because of their broad scope and widespread effects, which in turn control the other two categories, this research selected macroeconomic factors to analyse their impact on credit risk.

*2.3. Theme 3: Macroeconomic Determinants of Credit Risk*

In Table 1, we present a summary of the literature review on the macroeconomic variables.

**Table 1.** Review of macroeconomic variables.

| Variable Name | Definition | Abbreviation | Variable Type | Justification for Selection of Study Variables | Citation |
|---|---|---|---|---|---|
| Net national income | Amount of money generated within the nation's economy | NET_NATIO NAL_INCOME | Independent variable | It is the core revenue measure and denotes the source of income. It signifies a long-term economic growth indicator and the existing literature shows it as an NPL determinant. | [18,19] |
| National savings | Nation's wealth leading to investments | NATIONAL_SAVINGS | Independent variable | It indicates wealth growth and spending capacity. There is a scarcity of empirical studies that confirm national savings' impact on NPLs. Few authors have explored bank savings' impact on NPLs rather than the country's savings at the macroeconomic level. According to controversial Keynesians' "Paradox of Thrift" theory, which contends that when everyone starts to save more, aggregate demand decreases, there has been ambiguity about the impact of national savings on NPLs. | [10,15,20] |
| Employment rate | Percent of employed persons out of the total population | EMPLOYMENT_RATE | Independent variable | It is a unique variable that is present in both economic growth and the business cycle. Economists group normal employment, self-employment and entrepreneurship under employment, differentiated employment, and unemployment, according to the new classical school of thought. This research is one of the few of its kind that investigates the influence of the employment rate on NPLs as well as the impact of the unemployment rate, considering their individual significance. | [21,22] |
| Unemployment rate | Percent of unemployed persons out of total population | UNEMPLOYMENT _RATE | Independent variable | We chose this variable because it is a primary macroeconomic predictor of NPLs. | [10,13,15,23,24] |
| GDP quarter-to-quarter growth rate | Quarterly change rate in nation's real gross domestic product | GDP_QTQ_GROWTH _RATE | Independent variable | It is the core indicator of an economy's health. To deal with the current issue of stagflation in the UK (low GDP + high inflation), we chose to investigate this further. | [11,15,23,25,26] |

**Table 1.** *Cont.*

| Variable Name | Definition | Abbreviation | Variable Type | Justification for Selection of Study Variables | Citation |
| --- | --- | --- | --- | --- | --- |
| GBP-to-USD exchange rate | Conversion rate of GBP to USD | GBP_USD_EXCHANGE_RATE | Independent variable | A nation's power is seen to be reflected in the strength of its currency; therefore, several studies have identified the exchange rate as a critical predictor of NPLs around the world, and reveal that currency appreciation and depreciation have a significant impact on international trade borrowers' profitability. According to fewer studies, depreciation has a negative impact on NPLs, where devaluation has a greater impact on countries with large currency mismatches. On the other hand, depreciation stimulates export activity, which improves firms' financial conditions and enhances their capacity to pay. Thus, this work chooses to conduct more research to corroborate the confusing link between the exchange rate and NPLs from the UK's standpoint. | [24,27,28] |
| Total trade deficit | Country's import exceeds its exports | TOTAL_TRADE_DEFICIT | Independent variable | It is a crucial macroeconomic indicator of the business cycle, which also indicates supply and demand in industrialized global commerce. Despite the fact that it has long-term indirect effects on NPLs, few studies have been carried out to explore the impact of trad imbalances on bank credit risk. | [29] |
| Inflation rate | Overall increase in prices and increase in the cost of living | INFLATION_RATE | Independent variable | It demonstrates the buying power of money and is a primary macroeconomic predictor of NPLs. The existing literature contains inconsistent, contrary findings about the positive or negative impacts of inflation on NPLs. This highlights a gap and demands additional in-depth research. | [16,17,23,25] |
| National debt as percent of GDP | Ratio of country's public debt to its GDP | NATIONAL_DEBT_AS_PERCENT_GDP | Independent variable | It can create global and/or domestic market panic when it arises. Multiple empirical investigations present a high association pattern between two economic crises, where bank failures are typically preceded by uncontrolled national indebtedness. | [2,10,11] |

### 2.4. Theme 4: Credit Risk Predictive Models Using Macroeconomic Determinants

The fourth theme delves into business analytics, which is grouped into four unique categories (descriptive, predictive, diagnostic, and prescriptive analytics) based on their de-cision support system (DSS) application objective, specific levels of intelligence, complexity, and business value [30].

Financial researchers prefer supervised ML algorithms to deal with complex, structured financial data since they have faster execution, greater accuracy, and less-expensive deployment [31]. Thus, this work employs supervised ML methods. Classification is a strategy to forecast discrete (primarily binary) results. Classification ML algorithms cater to solve multiple finance business problems like risk management, strategic hedging, option pricing, and classifying bankruptcy [32], because these algorithms outperform traditional statistical models via supervised learning with structured data [33]. Logistic regression and ML-based algorithms can address the problem of high credit risk predictive modelling, as they have established their extensive applications across industries. In Table 2, we summarise the ML algorithms reviewed.

**Table 2.** Predictive modelling techniques.

| **Logistic Regression** | **Neural Network** | **Decision Tree** |
|---|---|---|
| Finance scholars utilize this statistical classification approach to elucidate intricate relationships among variables, gaining benefits in variable selection and coefficient shrinkage through cross-validation [34]. Logistic regression does not require a linear relationship between the response and predictor variable but the former must be categorical. The assumption of normal distribution may not always be applicable in real-world scenarios that can be characterized by non-linear data and correlated variables [35,36]. Consequently, this study also considers nonparametric models that do not rely on assumptions about data distribution. | Neural networks are becoming increasingly popular among scholars in the finance domain for credit risk evaluation because they outperform in statistical features like logistic regression and optimisation approaches [37]. The opposing strand of researchers is critical of their application as it is unstable, depends on the sample, and requires extensive computation and lengthy execution periods, which makes it difficult to conclude the optimal neural network [38]. The primary benefit is their strong generalisation ability. However, they are black-box models that are difficult for humans to interpret [39]. | The most popular ML technique for predicting credit risk and identifying financial fraud is the decision tree, which is a non-parametric and supervised learning technique [40]. One empirical investigation found that because decision trees are particularly sensitive to unbalanced data, they are the perfect choice for early credit risk warning [41], where taking preventative actions months in advance to avoid potential financial losses is essential [42]. Also, decision trees are explainable and easy to interpret compared to most conventional machine learning techniques, making them appealing to non-computing disciplines such as finance and economics. |

*2.5. Theme 5: Data Visualization of Credit Risk and Its Macroeconomic Determinants*

Data visualization is increasingly being used by banks to improve their DSS and is proving to be a valuable tool. Numerous studies recognise the growing appeal of data visualisation tools, highlighting their several advantages for DSSs, like real-time data analysis, multidimensional analysis, efficient insight portrayal, etc. [43,44]. To summarize the literature review, there are a few gaps and a scarcity of macroeconomic elements in the reviewed literature, which keep the research questions unanswered. Table 3 summarises the identified research gaps and comprehensive technical solutions to address those gaps.

**Table 3.** Summary of research gaps identified.

| **Research Gap** | **Proposed Solution** |
|---|---|
| There are US-focused NPL research studies that cover different scenarios: baseline (most likely scenario—low-credit-risk zone) or economically adverse (stress scenarios—high-credit-risk zone) [17]. | This study includes a comprehensive analysis of the UK's NPL data from 2005 to 2021 to cover various scenarios, such as baseline (most likely scenario—low-credit-risk zone) and economically adverse scenarios (stress scenarios—high-credit-risk zone). |
| Very few studies investigate national savings as the driver of credit risk, and those that do refer to data from savings banks and not the macroeconomic national savings data [10]. | This study examines the behaviour of the UK's national savings data from the credit risk perspective. |

**Table 3.** *Cont.*

| Research Gap | Proposed Solution |
|---|---|
| Existing studies do not consider a comprehensive outlook of employment status [21]. There is a need to cover both employment and the unemployment rate simultaneously against credit risk. | This study is unique in that it examines the impact of the employment rate and unemployment rate on NPLs separately and treats the employment rate as a distinct macroeconomic indicator. |
| There is a contradictory view about the UK currency exchange rate's impact on NPLs [27], which needs detailed investigation. | This study validates the conflicting association between the UK's currency exchange rate and NPLs. |
| The impact of trade deficit on credit risk has not received much attention [29]; thus, there is a need to examine the effects of the UK's trade deficits on NPLs. | This study extensively analyses the UK's trade deficit data and NPL association. |
| The literature investigating the link between NPLs and the inflation rate has inconsistent and contrary findings about the link [16,17,23,25], which clearly demands additional in-depth investigations. | This study extensively analyses the UK's inflationary data and NPL link. |
| There is another gap which reveals that the majority of studies only concentrate on the definition of risk (potential loss value or uncertainty of outcome) [9]. | This study integrates a binary target variable while retaining the original numeric target variable to cater both aspects of risk, estimating real credit risk value and the probability. |
| While there are numerous studies, as examined in the literature review section, on econometrics and big data analytics, very few address problem solution by combining the knowledge of banking, finance industry expertise, and advanced analytics. | This study implements advanced analytics such as predictive, descriptive, diagnostic, trend analysis, and the correlation of each study variable from the banking and finance industry. This research not only supplements but mitigates the strengths and weaknesses of both targeted domains. Thus, this research delivers an excellent blend of advanced analytics and banking–finance domain expertise. |

## 3. Methodology

This section presents the methods adopted in this study. Firstly, we employed the cross-industry standard data mining (CRISP-DM) framework for our analysis. Advanced analytical tools like the Analytics Software & Solutions (SAS) Enterprise Guide, Miner 9.4, and Tableau 2022 were employed. We used secondary data, where no human participation was involved in the collection of the data; therefore, ethical approval was not required [45]. The data were collected for the timeframe of 2005–2021. Firstly, we used data on the UK'S NPLs (frequency: quarterly) from the World Bank [46] and macroeconomic data (frequency: quarterly) from the UK's Office for National Statistics (ONS) [47].

### 3.1. Variable Selection Technique

The selection of variables is a crucial prerequisite step to include important variables in the model. We used a supervised learning strategy, named the information value (IV), to estimate the strength of the relationship between the independent and dependent variables. The higher the IV value, the greater the predictability. Table 4 depicts most of the shortlisted variables have extremely high predictive potential, except for TOTAL_TRADE_DEFICIT.

**Table 4.** Information value (IV) of variables.

| Variable | Information Value |
|---|---|
| INFLATION_RATE | 3.3035 |
| NET_NATIONAL_INCOME | 2.1769 |
| EMPLOYMENT_RATE | 1.9857 |
| GBP_USD_EXCHANGE_RATE | 1.5367 |
| UNEMPLOYMENT_RATE | 1.5291 |
| NATIONAL_DEBT_AS_PERCENT_GDP | 1.1663 |
| NATIONAL_SAVINGS | 0.8671 |
| GDP_QTQ_GROWTH_RATE | 0.8505 |
| TOTAL_TRADE_DEFICIT | 0.7545 |

Furthermore, we employed various criteria such as the adjusted R-square, mean squared error (MSE), and cross-validation prediction sum of squares (Cp) to identify more meaningful significance, i.e., the most parsimonious variables amongst the dataset. The execution approach is to select the combination of higher adjusted R-square values with the lowest Cp value and MSE. This study identifies the variables listed below as the most parsimonious ones, with the highest adjusted R-square value of 0.9164 with the lowest MSE and Cp values of 0.11112 and 5.0367, respectively:

"NATIONAL_SAVINGS, UNEMPLOYMENT_RATE, INFLATION_RATE, NATIONAL _DEBT_AS_PERCENT_GDP, GBP_USD_EXCHANGE_RATE".

*3.2. Data Processing*

Data pre-processing has a substantial impact on the predictive modelling quality. Thus, the subsequent sections discuss the data processing approach employed.

### 3.2.1. Removal of Duplicate Records

Since we collected macroeconomic variable data from multiple-source Excel files, we used the DISTINCT option for each file import to exclude duplicate records.

### 3.2.2. Handling of Missing Data

Despite the fact that the highly structured financial dataset contains no missing values, we advocated missing data imputation using the StatExplore node as best practise to enhance the statistical power.

### 3.2.3. Variable Renaming, Uniform Formatting, and Sorting

To improve the accuracy of the predictive modelling, we implemented cosmetic improvements such as uniform formatting, sorting, etc. [48].

### 3.2.4. Dealing with Outliers

We cautiously employed a "knowledge-based outlier analysis" approach to explore the outlier's beneficial features (the best or worst instance of the dataset), which has various practical applications like financial fraud detection, medical procedure test analysis, and scientific advancements [49]. A filter node with the "Extreme Percentile" option was used along with another best practise for outlier analysis, which is the examination of measures like leverage, deleted residuals, and the covariance ratio. In this way, raw data are processed with best practises to enhance predictive modelling.

*3.3. Data Transformation*

This section highlights critical steps to transform processed data into meaningful insights.

### 3.3.1. Append Data

The multiple pre-processed datasets are merged into a single dataset for further data analysis and predictive modelling.

### 3.3.2. Create New Binary Target Variable

To avert the loss of meaningful data, we added a new binary target variable, HIGH _CREDIT_RISK, to the original numeric target variable BANK_NLP_TO_GROSS_LOAN _PERCENT, which enables comprehensive descriptive, diagnostic, and predictive data analytics.

**4. Results**

The most valuable asset, according to British mathematician Clive Humby, is "the new oil" [50]. Thus, this section explores the underlying data in a variety of ways to deliver data-driven DSSs to targeted beneficiaries.

*4.1. Trend Analysis*

Targeted beneficiaries benefit from trend analysis as it reveals the trajectory of macroeconomic variables and credit risk over time by examining the underlying financial data pattern across the horizontal time axis and attempting to forecast future values based on historical data [51]. It is recommended to implement trend analysis prior to predictive modelling because it provides rudimentary yet insightful information about the underlying dataset [52]. We employed a horizontal trend analysis using Tableau to forecast future trends, where trend lines trace the movement of quantitative, continuous data. We opted for polynomial trend lines for variables with fluctuating underlying data and normal linear trend lines for variables with consistent underlying data, with significance at *p*-values < 0.05. We are aware of the predictive limitations of trend analysis, as it may not be suited for extreme or sudden changes [53]; thus, we included other analytical options, as mentioned in next the sections.

Except for the GBP_USD_EXCHANGE_RATE and TOTAL_TRADE_DEFICIT trend lines, all other trend lines reflect an upward trend at the tail of each graph and show the impact of the 2008 recession and the COVID-19 crisis, with a modest rising trend starting in 2019, as shown in Figure 1. The BOE identified inflation as the greatest significant risk in its report and encourages all UK institutions to undertake preventive measures [54]. According to the World Bank and the IMF, with a 256% increase in borrowing, global debt from both developed and emerging nations has risen to USD 226 trillion [55]. This analysis demonstrates that the research findings accurately depict real-world events.

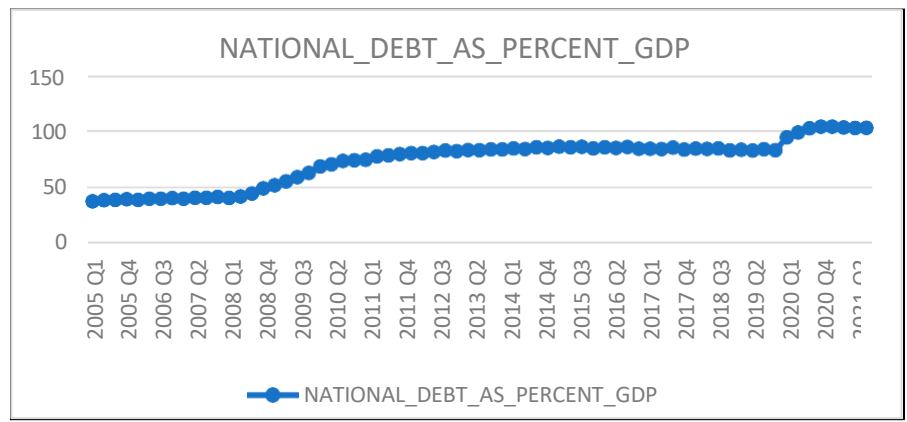

**Figure 1.** Trend analysis.

We performed a similar analysis for the remaining variables.

*4.2. Multidimensional Analysis*

The MultiPlot node provides multidimensional data visualisation. It explores underlying data graphically to understand data distributions and associations amongst variables. This research is notable for providing a comprehensive assessment of credit risk, such as the high or low credit risk probability for a given macroeconomic variable along with the real credit risk value. There is an approximate 7–10% chance of high credit risk when the national debt increases and the mean value of BANK_NLP_TO_GROSS_LOAN_PERCENT for the highest observation is around four. This research depicts negative links between GDP_QTQ_GROWTH_RATE and GBP_USD_EXCHANGE_RATE against BANK_NLP_TO_GROSS_LOAN_PERCENT (mean) and HIGH_CREDIT_RISK, as depicted in Figure 2. It indicates that a slowdown in GDP and the currency exchange rate can result in a rise in banks' credit risk. According to this study, there is an approximate 15% chance of high credit risk when the exchange rate depreciates at 1.56, and the mean value of BANK_NLP_TO_GROSS_LOAN_PERCENT for this observation is around three. Similarly, there is approximate 20% chance of high credit risk when the UK's quarterly GDP growth decreases when GDP_QTQ_GROWTH_RATE = 3, and the mean value of BANK_NLP_TO_GROSS_LOAN_PERCENT for this observation is around 2.5. This research shows the direct impact of INFLATION_RATE and NATIONAL_DEBT_AS _PERCENT_GDP on BANK_NLP_TO_GROSS_LOAN_PERCENT (mean) and HIGH _CREDIT_RISK, as shown in Figure 3. It indicates that a higher level of inflation and national debt increases banks' credit risk. There is strong evidence that national debt has a significant impact on the economy and increases the risk of default for banks. According to this research, there is an approximate 7–8% chance of high credit risk when the inflation rate increases consecutively in more than three quarters, and the mean value of BANK_NLP_TO_GROSS_LOAN_PERCENT for this observation is around 3.5.

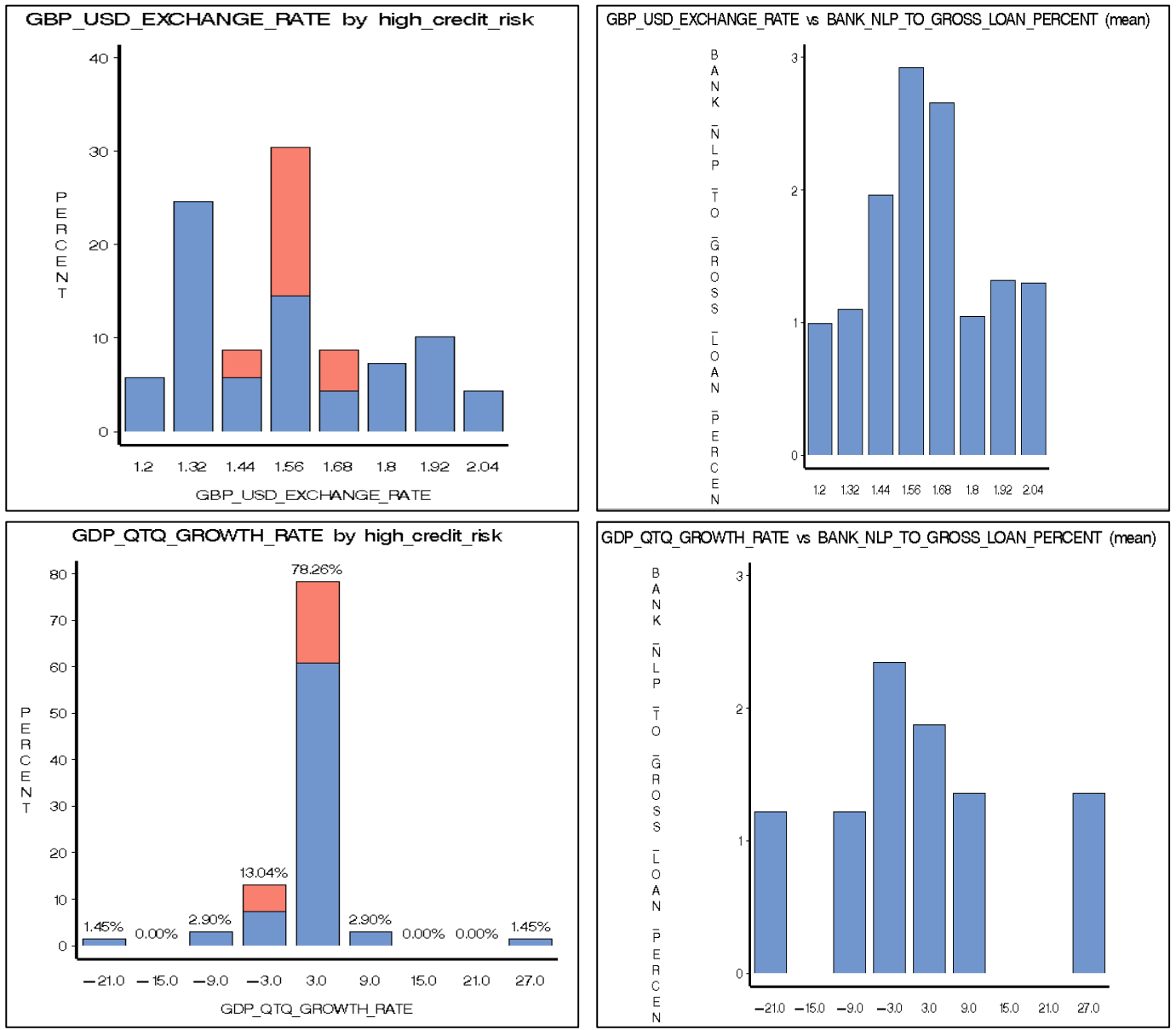

**Figure 2.** Multidimensional analysis 1.

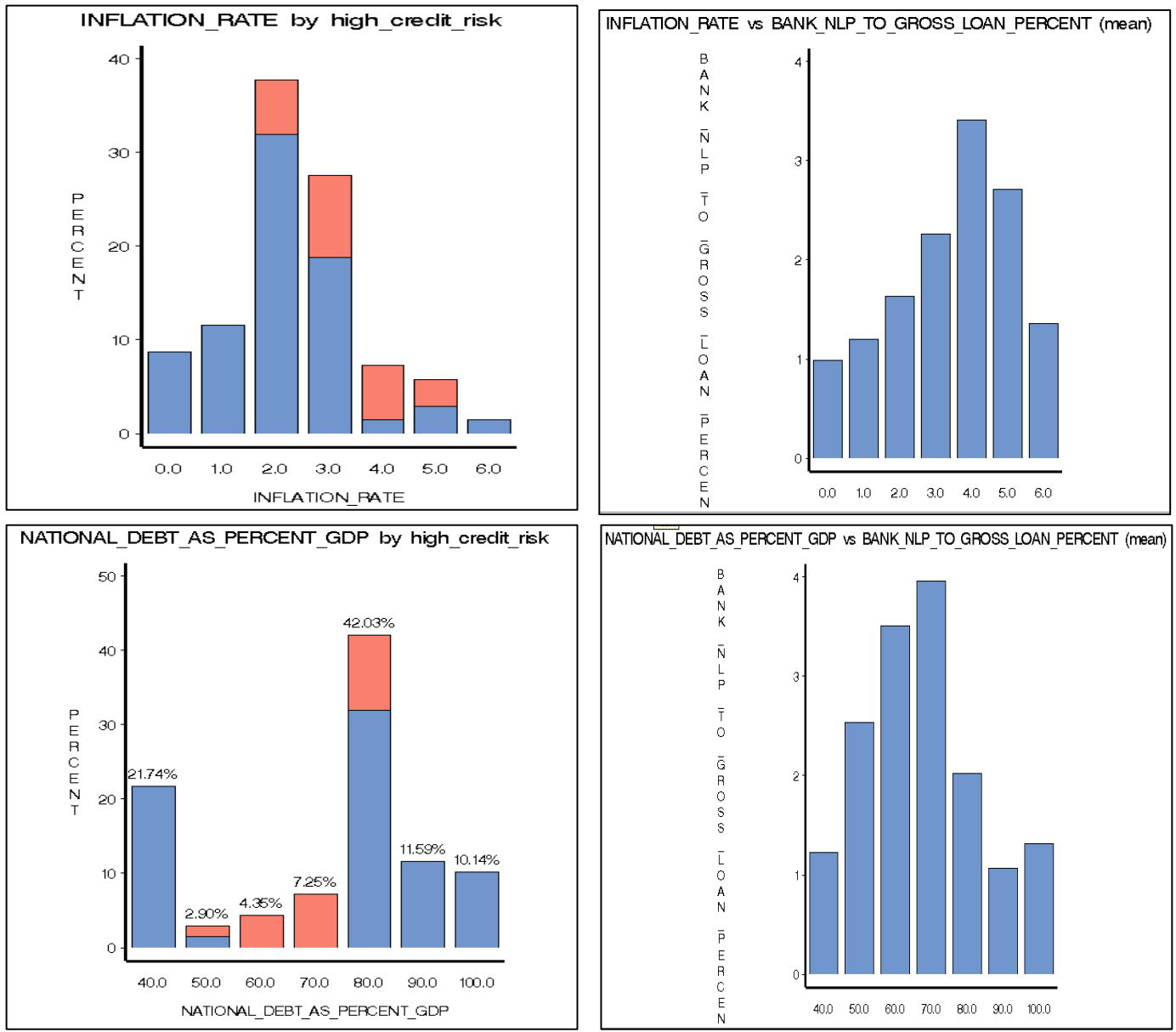

**Figure 3.** Multidimensional analysis 2.

We performed a similar analysis for the remaining variables.

*4.3. Descriptive Analysis*

Descriptive analysis is ideal for quantitative data that explains, depicts, and summarises constructive data points to analyse underlying data. Descriptive statistical analysis indicates meaningful data depiction with better interpretation, because raw data are challenging to visualize and comprehend [56]. Although descriptive analysis has been used in multiple finance studies, they might have explored statistical data and their business inferences in more depth [57]. This research delivers extensive business insights from descriptive analysis as value-added business intelligence to its beneficiaries. One noteworthy conclusion is evidence of the UK's stagflation, which clearly depicts slowed and troubled GDP growth with rising inflation [58]. The lowest GDP value is negative (−5.6), and the mean GDP value is extremely low (0.21875) for both high- and low-credit-risk zones, demonstrating that adverse financial conditions like the 2008 recession and COVID-19 have consistently and severely impacted the UK's economy. The higher standard deviation of GDP (2.8265) over the agreed-upon timescale shows greater volatility and a low consistency.

Inflation is another component of stagflation, which exhibits a higher coefficient of variation (29.2038), implying higher variability around the mean, as shown in Figure 4. This research expands upon a basic descriptive analysis of five-number summaries (lowest, median, quartile three, and maximum). The mean (3.2) and median values (3.2) of the inflation rate are approximately the same for both low and high credit risk. High credit risk means that the relative inflation rate varies on an approximately similar scale and is consistent as that of low credit risk, indicating a normal dispersion of data. The inflation rate remains at slightly higher levels, while the low-credit-risk zone's inflation rate remains at lower levels. The minimum range of the (lower whisker tail) inflation rate from the high-credit-risk zone exceeds the minimum values of the inflation rate of the low-credit-risk zone, the same as that for the maximum range, as shown in Figure 5. This indicates that the inflation rate is a potential determinant of credit risk. The high credit risk's mean relative unemployment rate is more consistent than the low credit risk's mean relative unemployment rate. More than 3% of the unemployment rate from the high credit risk shows higher values than that from low credit risk. The unemployment rate is more consistent in the high-credit-risk zone and remains at higher levels, while the low-credit-risk zone's unemployment rate varies, especially at lower levels. The minimum and maximum range (lower and higher whisker tail) of the unemployment rate from the high-credit-risk zone exceeds the minimum and maximum values of the unemployment rate of the low-credit-risk zone. Box plots also convey information about distribution shapes, specifically the skewness of the distribution. The majority of the unemployment rate falls below the median line for low credit risk, as shown in Figure 5. It indicates that the unemployment rate in the low-credit-risk zone has a slightly positive skewed distribution. We conducted a similar analysis for the remaining variables. The fundamental disadvantage of descriptive analytics is that it only offers retrospective analysis without attempting to uncover the causes or anticipate the future [59]. Thus, in the following sections, we explore diagnostic and predictive analysis.

| HIGH_CREDIT_RISK | N Obs | Label | Mean | Std Dev | Std Error | Minimum | Maximum | Median | Coeff of Variation |
|---|---|---|---|---|---|---|---|---|---|
| 1 | 16 | NET_NATIONAL_INCOME | 347893.63 | 14961.00 | 3740.25 | 317847.00 | 370204.00 | 350692.50 | 4.30 |
| | | NATIONAL_SAVINGS | 99522.13 | 2582.01 | 645.5034849 | 94555.00 | 102664.00 | 100386.00 | 2.59 |
| | | EMPLOYMENT_RATE | 70.6625000 | 0.4364631 | 0.1091158 | 70.1000000 | 71.7000000 | 70.5500000 | 0.62 |
| | | UNEMPLOYMENT_RATE | 7.9000000 | 0.2851900 | 0.0712975 | 7.1000000 | 8.4000000 | 7.9000000 | 3.61 |
| | | GDP_QTQ_GROWTH_RATE | 0.2187500 | 2.8265925 | 0.7066481 | -5.6000000 | 2.9000000 | 1.1500000 | 1292.16 |
| | | GBP_USD_EXCHANGE_RATE | 1.5748875 | 0.0529471 | 0.0132368 | 1.4346000 | 1.6411000 | 1.5808500 | 3.36 |
| | | TOTAL_TRADE_DEFICIT | -5696.38 | 2081.97 | 520.4918257 | -9167.00 | -1491.00 | -5633.00 | -36.54 |
| | | INFLATION_RATE | 3.2000000 | 0.9345231 | 0.2336308 | 1.5000000 | 4.7000000 | 3.2000000 | 29.20 |
| | | NATIONAL_DEBT_AS_PERCENT_GDP | 71.8625000 | 9.7748572 | 2.4437143 | 51.8000000 | 82.7000000 | 74.2000000 | 13.60 |
| | | BANK_NLP_TO_GROSS_LOAN_PERCENT | 3.7529108 | 0.2140782 | 0.0535195 | 3.5086570 | 3.9618682 | 3.7705589 | 5.70 |
| 0 | 53 | NET_NATIONAL_INCOME | 400497.77 | 60341.51 | 8288.54 | 305497.00 | 519131.00 | 401315.00 | 15.06 |
| | | NATIONAL_SAVINGS | 129764.64 | 45577.45 | 6260.54 | 67825.00 | 217367.00 | 131000.00 | 35.12 |
| | | EMPLOYMENT_RATE | 73.9264151 | 1.4243961 | 0.1956558 | 71.2000000 | 76.5000000 | 73.8000000 | 1.92 |
| | | UNEMPLOYMENT_RATE | 5.1264151 | 0.9952099 | 0.1367026 | 3.8000000 | 7.8000000 | 5.1000000 | 19.41 |
| | | GDP_QTQ_GROWTH_RATE | 1.5226415 | 5.1939551 | 0.7134446 | -21.1000000 | 24.5000000 | 2.2000000 | 341.11 |
| | | GBP_USD_EXCHANGE_RATE | 1.5485396 | 0.2556874 | 0.0351214 | 1.2326000 | 2.0444000 | 1.5171000 | 16.51 |
| | | TOTAL_TRADE_DEFICIT | -6621.83 | 5898.45 | 810.2144006 | -23152.00 | 17596.00 | -7148.00 | -89.07 |
| | | INFLATION_RATE | 2.0358491 | 1.2439228 | 0.1708659 | 0 | 6.2000000 | 2.1000000 | 61.10 |
| | | NATIONAL_DEBT_AS_PERCENT_GDP | 73.6962264 | 22.7812757 | 3.1292489 | 37.6000000 | 103.8000000 | 83.8000000 | 30.91 |
| | | BANK_NLP_TO_GROSS_LOAN_PERCENT | 1.3205700 | 0.5707002 | 0.0783917 | 0.7346597 | 3.1117417 | 1.1278600 | 43.21 |

**Figure 4.** Descriptive analysis 1.

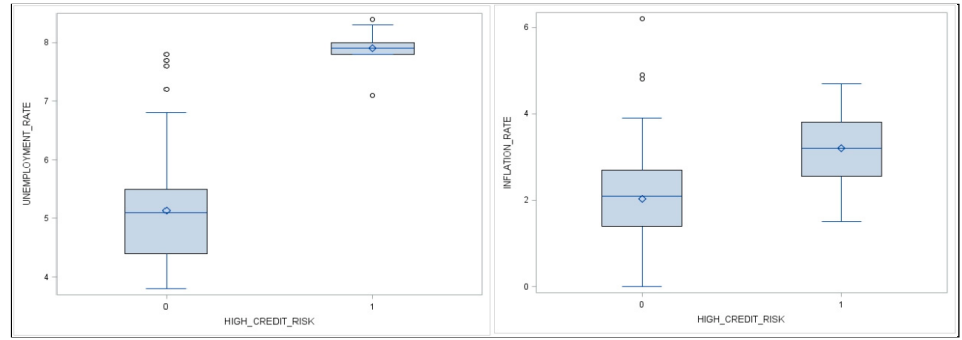

**Figure 5.** Descriptive analysis 2.

We performed a similar analysis for the remaining variables.

*4.4. Distribution Analysis*

Histograms, probability, and quantile–quantile (QQ) plots were utilized for the distribution analysis of quantitative data. The important aspect of distribution analysis is its implicit use in statistical testing (for example, multicollinearity uses the F-test and T-test, and decision tree models employ the Chi-test for validation).

The analysis validates the normal distribution of all variables, where most of the histograms are unimodal (one data peak) or bimodal (two data peaks). While the majority of histograms are symmetric, the UNEMPLOYMENT_RATE histogram displays slight positive skewness, as depicted in Figure 6.

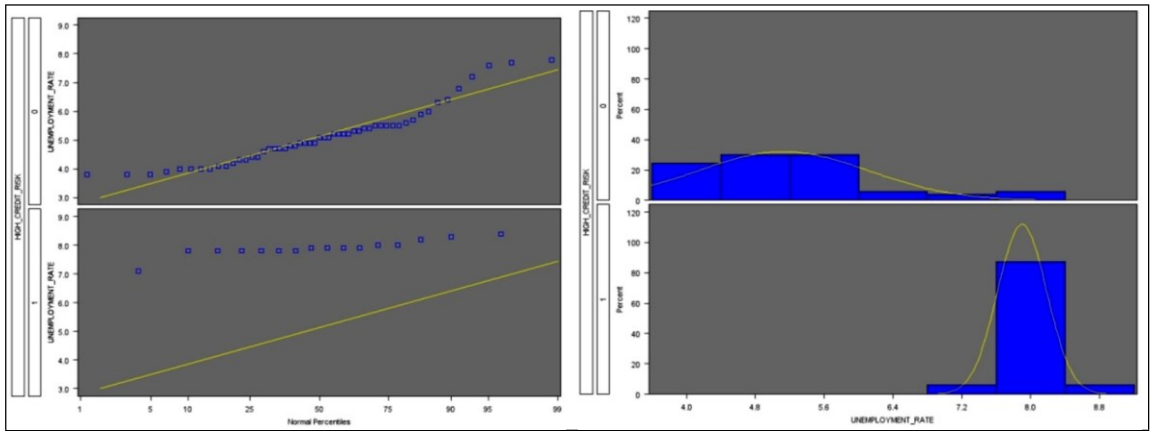

**Figure 6.** Distribution analysis of UNEMPLOYMENT_RATE.

We performed a similar analysis for the remaining variables.

*4.5. Multicollinearity*

Examining multicollinearity prior to predictive data modelling is suggested as the best practice. In this study, we utilised the variance inflation factor (VIF), which is formulated below.

$$\text{VIF}_j = \frac{1}{1 - R_j^2} \quad (1)$$

where j is number of variables and $R^2$ is the coefficient of determination.

The key indication of multicollinearity in the dataset is confirmed by VIF values which are greater than 10 and validated by the considerably higher adjusted $R^2$ = 0.9151. The parameter estimate in Table 5 illustrates that variables with VIF values less than 10 are dependent on those with VIF values more than 10, thus adequately proving substantial collinearities. Furthermore, we performed correlation between the variables as reported in Table 6.

Table 5. VIF values of variables.

| Variable | VIF |
| --- | --- |
| NET_NATIONAL_INCOME | 29.77 |
| NATIONAL_SAVINGS | 14.35 |
| EMPLOYMENT_RATE | 88.08 |
| UNEMPLOYMENT_RATE | 53.91 |
| GDP_QTQ_GROWTH_RATE | 1.88 |
| GBP_USD_EXCHANGE_RATE | 4.61 |
| TOTAL_TRADE_DEFICIT | 1.45 |
| INFLATION_RATE | 1.65 |
| NATIONAL_DEBT_AS_PERCENT_GDP | 11.7 |

Table 6. Correlation coefficient values.

| Correlation Coefficient Range | Interpretation | Correlations Pairs with Correlation Coefficient |
| --- | --- | --- |
| 0.9 to 1.0 | positive and a very strong correlation | NET_NATIONAL_INCOME—NATIONAL_SAVINGS (0.9289) |
| 0.3 to 0 | positive and a very weak (low) or negligible correlation | GDP_QTQ_GROWTH_RATE—GBP_USD_EXCHANGE_RATE (0.1464) |
| −1 to −0.9 | negative and a very strong correlation | EMPLOYMENT_RATE—UNEMPLOYMENT_RATE (—0.9466) |

We further conducted a diagnostic collinearity analysis to derive statistical inference, which analyses condition indices to identify which independent variables are most closely associated with each other. Independent variables like EMPLOYMENT_RATE (0.99814) and UNEMPLOYMENT_RATE (0.90645) show reasonably large loadings (coefficients), with a close-to-zero Eigenvalue for row number 10 with the highest condition index of 1528. This demonstrates the co-linear nature of EMPLOYMENT_RATE and UNEMPLOYMENT_RATE; nonetheless, this may have an effect on the prediction of the dependent variable BANK_NLP_TO_GROSS_LOAN_PERCENT. As a result, while building the predictive model, SAS chooses either of them based on the highest loading and correlation coefficient by default.

*4.6. Diagnostic Analysis*

Diagnostic analysis helps to determine the source of a significant correlation. The magnitude and direction of multivariate data distributions in multidimensional space are represented by the covariance matrix values. For example, "Why is the UK's NATIONAL_DEBT_AS PERCENT_GDP mounting in the provided timeframe, and what might be the root cause for the same?" The negative covariance coefficient from Table 7 shows that as the UK's currency depreciates over time, the NATIONAL_DEBT_AS PERCENT_GDP rises.

Table 7. Covariance matrix.

| Covariance Matrix | |
| --- | --- |
| | NATIONAL_DEBT_AS_PERCENT_GDP |
| GBP_USD_EXCHANGE_RATE | −3.772 |

This research illustrates the previous example of diagnostic analysis from the dataset and uses a similar diagnostic approach for the remaining variables. In this way, we conduct multidimensional data analysis to generate multiple significant data insights, prior to predictive modelling with improved data understanding.

*4.7. Discussion*

The goal is to increase the understanding of underlying macroeconomic causes of credit risk and predict the macroeconomic variables to which credit risk is most sensitive with accuracy (high- and low-credit-risk zones). This will enable beneficiaries to deploy control mechanisms to prevent projected credit losses and maintain adequate reserves to comply with UK regulatory norms [4]. This section discusses the credit risk predictive model's development and comparison to select the most accurate model using measures like the confusion matrix and receiver operating characteristic curve (ROC). The input parameters for all predictive models (logistic regression models, neural network, and decision tree) are NET_NATIONAL_INCOME, NATIONAL_SAVINGS, EMPLOYMENT_RATE, UNEMPLOYMENT_RATE, GDP_QTQ_GROWTH_RATE, GBP_USD_EXCHANGE_RATE, TOTAL_TRADE_DEFICIT, INFLATION_RATE, and NATIONAL_DEBT_AS_PERCENT_GDP, and the target parameter is an SAS variable named HIGH_CREDIT_RISK based on BANK_NLP_TO_GROSS_LOAN_PERCENT.

4.7.1. Logistic Regression

We implemented stepwise, forward, and backward logistic regression algorithms by distinctively choosing the LOGIT function (apt for binary target variables for prediction robustness) over the PROBIT function, as well as cross-validation criteria for accurate predictability [60]. Table 8 illustrates the model equations of all three logistic regression models at the 95% significance level. As a consequence of the implementation of cross-validation and stratified data partition, the test results indicate consistent, minimum fluctuation across different datasets.

**Table 8.** Regression model equations.

| Model Description | Model Equation |
| --- | --- |
| Backward_Regression | HIGH_CREDIT_RISK = −40.1049 + 1.2648 (INFLATION_RATE) − 0.3805 (NATIONAL_DEBT_PERCENT_GDP) + 0.000514 (NATIONAL_SAVINGS) − 0.00021 (NATIONAL_INCOME) − 0.00117 (TOTAL_TRADE_DEFICIT) + 10.7913 (UNEMPLOYMENT_RATE) |
| Stepwise_Regression | HIGH_CREDIT_RISK = −34.24 + 4.6830 (UNEMPLOYMENT_RATE) |
| Forward_Regression | HIGH_CREDIT_RISK = −34.24 + 4.6830 (UNEMPLOYMENT_RATE) |

The findings of all three regression model equations reinforce the existing literature on macroeconomic factors' impact on credit risk. Conclusively, regression models demonstrate a positive link between credit risk and inflation [17], the unemployment rate [13], and national savings [20], as well as a negative link with the UK's national debt [2], total trade deficit [29], and national income [19].

4.7.2. Neural Network

We employed a neural network with a multilayer perceptron algorithm, with model selection criterion being "average error" to minimize the average error in the validation dataset. We explored the weight plot and understood the variable importance. The weight indicators ("+", "−") imply equivalent (positive, negative) associations of dependent (NPLs) and all independent macroeconomic variables, as shown in Table 9. The sign of the weight for all variables matches with link interpretations derived from logistic regression models. Remarkably, the "Paradox of Thrift" is confirmed by both the backward logistic regression model and the neural network.

Table 9. Weight of variable by neural network.

| Variable | Weight |
|---|---|
| UNEMPLOYMENT_RATE | 1.285231083 |
| NATIONAL_SAVINGS | 0.732948287 |
| INFLATION_RATE | 0.237606653 |
| TOTAL_TRADE_DEFICIT | −0.01663233 |
| GDP_QTQ_GROWTH_RATE | −0.03683999 |
| NET_NATIONAL_INCOME | −0.28950698 |
| NATIONAL_DEBT_AS_PERCENT_GDP | −0.95127149 |
| GBP_USD_EXCHANGE_RATE | −1.48455938 |
| EMPLOYMENT_RATE | −1.51530151 |

4.7.3. Decision Tree

We implemented a multi-split decision tree based on an algorithm that uses variance as the interval target criterion, ProbChisq as the nominal target criterion, and the average squared error as an assessment metric for sub-trees. The variance splitting criteria ensure stable predictability with few deviations [61]. This approach was adapted from data-driven observations of descriptive analysis, where the variance and standard deviation of independent variables were influenced by binary target variables. Thus, this study establishes the value of performing data analysis before predictive analysis.

The decision tree highlighted unemployment and inflation in the UK as the strongest determinants of credit risk as evidenced in Table 10, with the lowest average squared error and consistent testing results across different datasets, as depicted in Figure 7.

Table 10. Decision tree output analysis.

| Interpretation of Decision Tree |
|---|
| 1. If the unemployment rate in the UK exceeds 7.7, then there is a 100% chance of high credit risk—represented by 1. |
| 2. If the UK's unemployment rate is less than 7.7, which implies that it is not unmanageable, then there is around a 4% chance of high credit risk—represented by 1. |
| 3. If the UK's unemployment rate is less than 7.7 but the quarterly inflation rate exceeds 2.9, then there is a 20% chance of high credit risk—represented by 1. |
| 4. If the inflation rate in the UK remains less than 2.9, then there is no chance of high credit risk. |

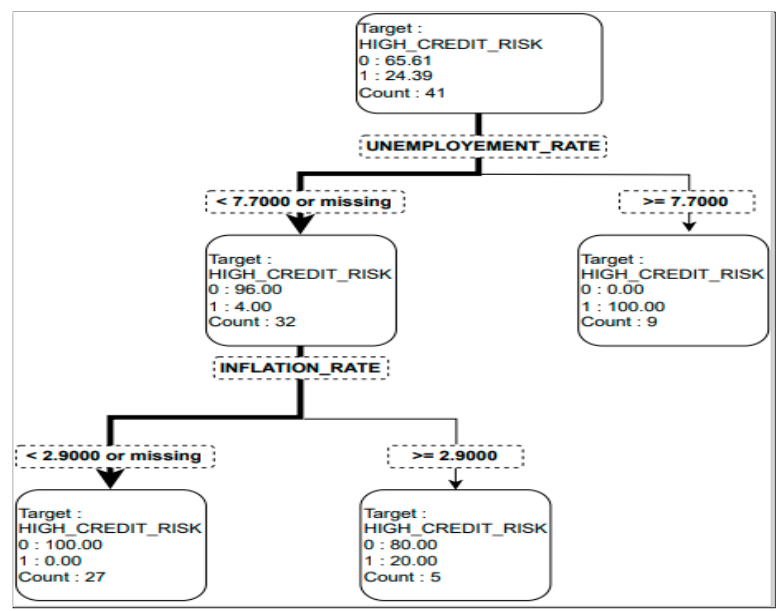

**Figure 7.** Decision tree.

4.7.4. Predictive Models' Comparison

Based on the validation dataset's average squared error, misclassification rate, and ROC index, we compared the developed models using the model comparison node and evaluated each model's scores using the score node.

The decision tree was selected as the most accurate and best-fitting model for the underlying dataset, which is consistent with findings of former studies [42]. Decision trees perform well in both balanced and unbalanced datasets, as seen in the generated model score box plot in Figure 8, which shows that its mean and median scores are similar and the most balanced.

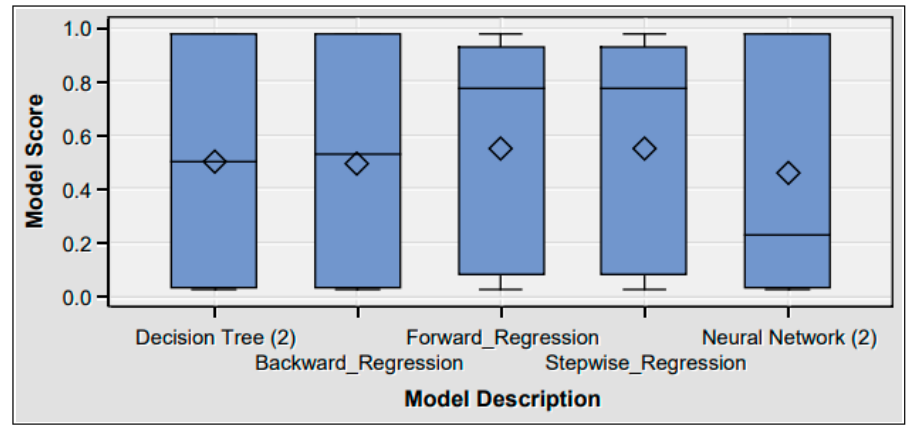

**Figure 8.** Model score.

The model comparison node generates the confusion matrix. We used five performance metrics to evaluate the efficacy of the predictive model as shown in Table 11 and explained further in Table 12.

**Table 11.** Model comparison.

| Model Node | Data Model | Recall | Precision | Sensitivity | Specificity | Accuracy |
|---|---|---|---|---|---|---|
| Tree2 | Decision_Tree | 0.75 | 1 | 0.75 | 1 | 0.95 |
| Neural2 | Neural_Network | 0.5 | 0.67 | 0.5 | 0.93 | 0.85 |
| Reg2 | Stepwise_Regression | 0.8 | 0.8 | 0.8 | 0.93 | 0.90 |
| Reg3 | Forward_Regression | 0.8 | 0.8 | 0.8 | 0.93 | 0.90 |
| Reg4 | Backward_Regression | 0.25 | 0.5 | 0.25 | 0.93 | 0.80 |

**Table 12.** Performance-measure findings.

| Performance Measures | Interpretation |
|---|---|
| Precision | The decision tree has the greatest precision, suggesting the generation of more relevant results than irrelevant ones. |
| Recall | Among all models, stepwise and forward regression models exhibit high recall scores. |
| Accuracy | When compared to the other models, the decision tree has the highest accuracy of 95%. The accuracy of the backward regression model is lowest due to the large number of predictor variables in the model equation. Accuracy has one limitation to deliver the best results for balanced data [62]. Thus, we assess two more additional performance indicators: sensitivity and specificity. |
| Sensitivity | Forward and stepwise logistic regression models are more sensitive to outliers than the other models, making them less robust to extreme values than decision trees, which do not divide trees based on outliers [63]. |
| Specificity | Inflation rate and unemployment rate are the most specific to the high-credit-risk zone. |

SAS generates the ROC curve automatically to compare the performance of the predictive model in terms of sensitivity (true positive) and specificity (false positive). The decision tree model (blue line) outperformed the other predictive models, according to the validation dataset's ROC chart from Figure 9. A series of classification charts produced through the model comparison demonstrates that the decision tree is effective at correctly classifying HIGH_CREDIT_RISK values in the validation dataset, as shown in Figure 10.

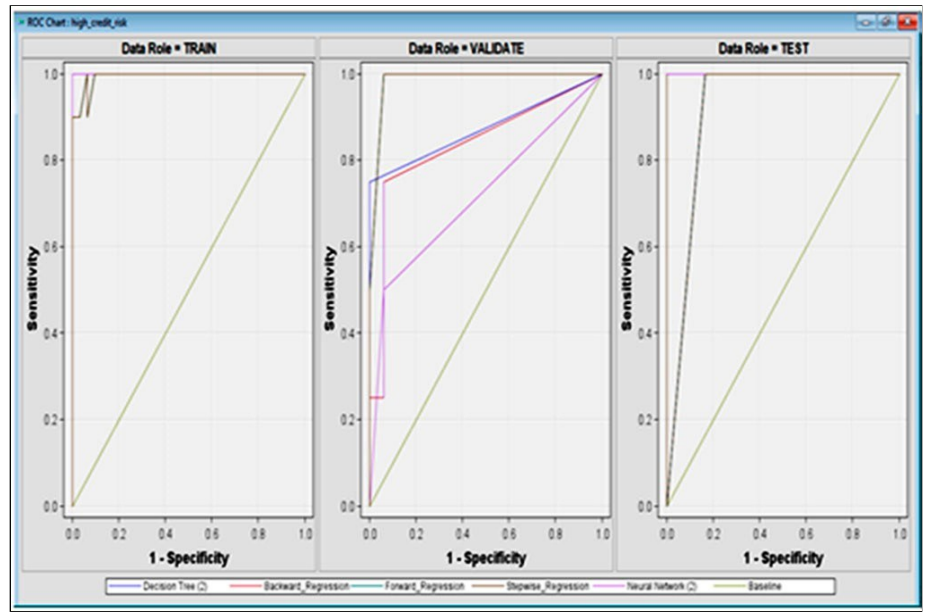

**Figure 9.** ROC chart.

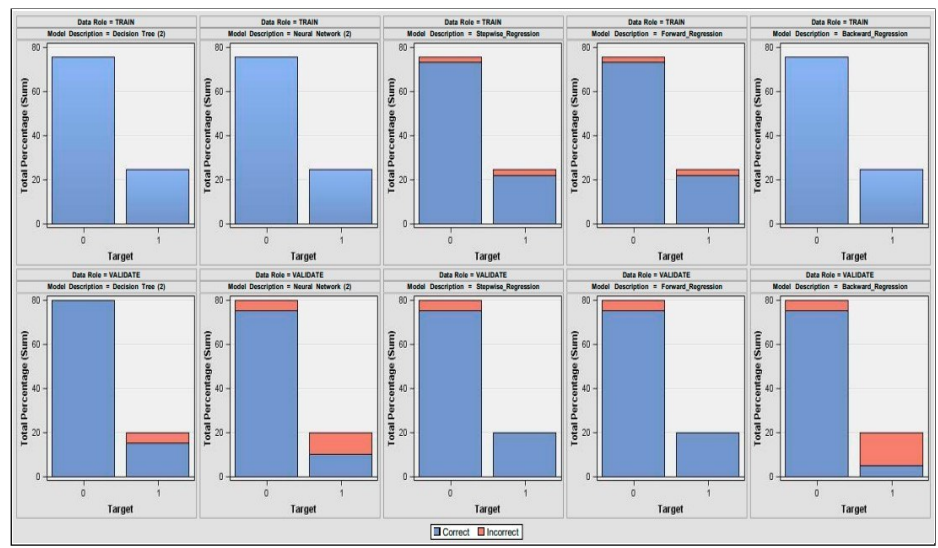

**Figure 10.** Classification chart.

## 5. Conclusions

This study aimed to investigate the macroeconomic determinants of the UK's banking credit risk from 2005 to 2021. To achieve this, we examined the impact of several macroeconomic factors on banks credit risk for the period of 2005–2021. Our findings distinctly establish NATIONAL_SAVINGS, UNEMPLOYMENT_RATE, INFLATION_RATE NATIONAL_DEBT_AS_PERCENT_GDP, and GBP_USD_EXCHANGE_RATE as the most parsimonious predictor variables, with the highest adjusted R-square value of 0.9164 and the lowest MSE and Cp values of 0.11112 and 5.0367, respectively. Furthermore, we ex-plored the trends of the UK's macroeconomic factors and banking credit risk from 2005 to 2021. The findings depict the trajectory of all macroeconomic factors which covered the baseline (most likely scenario—post 2008 recession) and economically adverse (stress scenarios—2008 recession, COVID-19 crisis) scenarios' slow recovery. All studied trend lines except for TOTAL_TRADE_DEFICIT and GBP_USD_EXCHANGE_RATE represent an upward trend at the tail of each graph. Our results showed plausible causes of the increase in the UK's national debt, such as currency depreciation and the increasing trade deficit over time, through a diagnostic analysis. This study also offers empirical evidence for the UK's stagflation through a graphical (box plot) descriptive analysis [58]. Our study found the variance-based multi-split decision tree to be the most accurate predictive model with consistent and robust predictability [42]. Most importantly, the model is interpretable, easy to implement, and reliable. Our findings suggest that unemployment and inflation are strong determinants of UK banking credit risk. This study empirically supported the findings of the existing literature on the influence of macroeconomic factors on credit risk, with a demonstration of a direct (positive) association between credit risk and inflation [17], the unemployment rate [13], and national savings [20]; an inverse (negative) relationship was established between credit risk and national debt [2], total trade deficit [29], and national income [19]. This paper significantly contributes to the empirical proof that the positive link between national savings and NPLs—that is, the "Paradox of Thrift"—is as follows: when savings rise, national wealth rises owing to a failure to spend money within the market, which slows the economy and impacts supply–demand (trade deficit), which in turn decreases GDP and enhances credit risk.

Interestingly, the research outcomes reflected the current state of the UK's macroeconomy and its influence on banks' credit risk. Even the BOE has warned about the latest fall in UK employment, which makes the BOE's goal of managing inflation even more difficult [37]. Thus, this paper empirically proved that the comprehensive advanced analytical findings are beneficial and informative for targeted beneficiaries (credit risk management teams) to conduct data-driven DSSs, to monitor macroeconomic factor thresholds, and to execute key measures to mitigate high credit risk.

This study offers the following technical best practises based on the aforementioned findings:

- To assure unaffected, reliable, accurate outcomes, we recommend conducting multicollinearity and trend analysis prior to commencing predictive data modelling;
- To improve predictive modelling execution, it is advisable to use missing data imputation, making aesthetic adjustments such as renaming, using consistent formatting for all variables in ascending order;
- This research employs yet another best practise, the extensive analysis of outliers, by analysing measures like leverage, deleted residuals, and the covariance ratio;
- For highly structured, normally distributed, quantitative data, the stratified technique of data partitioning is recommended, as it produces precise testing results with minimum variation compared to the simple random method with a comparable sample size.

The following enhancements may be included in future research.

Massive amounts of data processing causes computational complexities; thus, predictive analytics would benefit from distributed computing. Predictive analytics would improve credit risk "live" early warning systems by implementing live stream processing and complex event processing (CEP). The current predictive model is compatible with CEP technologies for collecting and processing live event data to detect patterns of high credit risk or vulnerable macroeconomic zones [64].

Although the UK's macroeconomic data for the last century are publicly available, banking credit risk (NPL) data are not. Therefore, our findings may be challenging to apply to other situations due to data availability over larger periods.

**Author Contributions:** Conceptualization, H.S. and A.A.; methodology H.S. and A.A.; software, A.A.; validation, H.S. and A.A.; formal analysis, A.A.; investigation, H.S. and A.A.; resources H.S., A.A., O.A. and B.O.; data curation, H.S., A.A., O.A. and B.O.; writing—original draft preparation, H.S., A.A.,O.A. and B.O.; writing—review and editing, H.S., A.A.,O.A. and B.O.; visualization, H.S., A.A., O.A. and B.O.; supervision, H.S.; project administration, B.O. All authors have read and agreed to the published version of the manuscript.

**Funding:** This research received no external funding.

**Institutional Review Board Statement:** Not applicable.

**Informed Consent Statement:** Not applicable.

**Data Availability Statement:** The dataset is open and publicly available at the Office for National Statistics (ons.gov.uk): https://www.ons.gov.uk (accessed on 1 May 2022).

**Conflicts of Interest:** The authors declare no conflict of interest.

## List of Abbreviations

| | |
|---|---|
| Analytics Software & Solutions | SAS |
| United Kingdom | UK |
| Bank of England | BOE |
| International Monetary Fund | IMF |
| Association for Computing Machinery | ACM |
| European Union | EU |
| Cross-industry standard Data Mining | CRISP-DM |
| General Data Protection Regulation | GDPR |
| UK Office for National Statistics | ONS |
| Decision support system | DSS |
| Machine learning | ML |
| Information value | IV |
| Non-performing assets | NPA |
| Ratio of capital adequacy | CAR |
| Non-performing loan | NPL |
| Gross Domestic Product | GDP |
| Great British Pound | GBP |
| United States Dollar | USD |
| Receiver Operating Characteristic Curve | ROC |
| Variance Inflation Factor | VIF |
| Complex event processing | CEP |
| Quantile–Quantile | QQ |